\documentclass[%
 reprint,
 noeprint,  %
superscriptaddress,
 amsmath,amssymb,
 aps,
floatfix,
]{revtex4-2}

\usepackage{graphicx}%
\usepackage{dcolumn}%
\usepackage{bm}%

\usepackage[caption=false]{subfig}

\usepackage{siunitx}
\DeclareSIUnit\ions{ions}
\DeclareSIUnit\ipsn{\ions \per \nm\squared}
\DeclareSIUnit\bar{bar}
\usepackage[version=4]{mhchem}
\usepackage{booktabs}

\usepackage[disable]{todonotes}

\usepackage{xcolor}
\definecolor{lightblue}{cmyk}{0.230667, 0.076, 0., 0.0706667}    %
\definecolor{darkred}{rgb}{0.717647,   0.086275,   0.086275}    %
\definecolor{darkred}{rgb}{0.85882   0.10196   0.10196}    %

\usepackage{hyperref}
\hypersetup{
    unicode=false,
    pdftoolbar=true,
    pdfmenubar=true,
    pdffitwindow=false,
    pdfstartview={FitH},
    pdftitle={Origin of performance enhancement of superconducting nanowire
single-photon detectors by He-ion irradiation},
    pdfauthor={Stefan Strohauer},
    pdfcreator={Latex with hyperref},
    pdfproducer={pdflatex},
    pdfkeywords={SSPD} {SNSPD} {He ion irradiation} {superconducting thin film} {transport measurement} {thermal conductance} {radiation damage} {He ion irradiation} {defect engineering},
    pdfnewwindow=true,
    pdfborder={0 0 0.5},
    colorlinks=false,
    linkcolor=red,
    citecolor=green,
    filecolor=magenta,
    urlcolor=cyan,
    allbordercolors={lightblue},
}
\usepackage[capitalize,noabbrev]{cleveref}

\newcommand{\dif}{\mathop{}\!\mathrm{d}}
\makeatletter
\newcommand{\spx}[1]{%
  \if\relax\detokenize{#1}\relax
    \expandafter\@gobble
  \else
    \expandafter\@firstofone
  \fi
  {^{#1}}%
}
\makeatother

\newcommand{\genericdel}[4]{%
  \ifcase#3\relax
  \ifx#1.\else#1\fi#4\ifx#2.\else#2\fi\or
  \bigl#1#4\bigr#2\or
  \Bigl#1#4\Bigr#2\or
  \biggl#1#4\biggr#2\or
  \Biggl#1#4\Biggr#2\else
  \left#1#4\right#2\fi
}

\makeatletter
\newcommand\thefontsize{The current font size is: \f@size pt}

\usepackage{xspace}    %

\newcommand{\etal}{\textit{et al.}\xspace} %

\newcommand{\sspd}{SNSPD\xspace}
\newcommand{\sspds}{SNSPDs\xspace}

\newcommand{\nbtin}{\ce{NbTiN}\xspace}

\newcommand{\siox}{\ce{SiO2}\xspace}
\newcommand{\Isw}{I_\mathrm{sw}\xspace}

\newcommand{\Ic}{I_\mathrm{c}\xspace}
\newcommand{\Rsheet}{R_\mathrm{sheet}\xspace}
\newcommand{\Tc}{T_\mathrm{c}\xspace}

\newcommand{\csub}{thermal conductance\xspace}
\newcommand{\csublong}{thermal conductance between the superconducting thin film and the substrate\xspace}

\begin{document}

\title{Origin of performance enhancement of superconducting nanowire single-photon detectors by He-ion irradiation}

\author{Stefan Strohauer}
\thanks{These two authors contributed equally}
\email{stefan.strohauer@tum.de}
\affiliation{Walter Schottky Institute, Technical University of Munich,
85748 Garching, Germany
}
\affiliation{TUM School of Natural Sciences, Technical University of Munich,
85748 Garching, Germany}
\author{Fabian Wietschorke}
\thanks{These two authors contributed equally}
\email{stefan.strohauer@tum.de}
\affiliation{Walter Schottky Institute, Technical University of Munich,
85748 Garching, Germany
}
\affiliation{TUM School of Computation, Information and Technology, Technical University of Munich,
80333 Munich, Germany
}
\author{Markus Döblinger}
\affiliation{Ludwig Maximilian University of Munich,
Faculty for Chemistry and Pharmacy, 81377 Munich, Germany
}
\author{Christian Schmid}
\affiliation{Walter Schottky Institute, Technical University of Munich,
85748 Garching, Germany
}
\affiliation{TUM School of Computation, Information and Technology, Technical University of Munich,
80333 Munich, Germany
}
\author{Stefanie Grotowski}
\affiliation{Walter Schottky Institute, Technical University of Munich,
85748 Garching, Germany
}
\affiliation{TUM School of Natural Sciences, Technical University of Munich,
85748 Garching, Germany}
\author{Lucio Zugliani}
\affiliation{Walter Schottky Institute, Technical University of Munich,
85748 Garching, Germany
}
\affiliation{TUM School of Computation, Information and Technology, Technical University of Munich,
80333 Munich, Germany
}
\author{Björn Jonas}
\affiliation{Walter Schottky Institute, Technical University of Munich,
85748 Garching, Germany
}
\affiliation{TUM School of Computation, Information and Technology, Technical University of Munich,
80333 Munich, Germany
}
\author{Kai Müller}
\affiliation{Walter Schottky Institute, Technical University of Munich,
85748 Garching, Germany
}
\affiliation{TUM School of Computation, Information and Technology, Technical University of Munich,
80333 Munich, Germany
}
\affiliation{Munich Center for Quantum Science and Technology (MCQST),
80799 Munich, Germany}
\author{Jonathan J. Finley}
\email{jj.finley@tum.de}
\affiliation{Walter Schottky Institute, Technical University of Munich,
85748 Garching, Germany
}
\affiliation{TUM School of Natural Sciences, Technical University of Munich,
85748 Garching, Germany}
\affiliation{Munich Center for Quantum Science and Technology (MCQST),
80799 Munich, Germany}

\date{January 21, 2025}

\begin{abstract}

Superconducting nanowire single-photon detectors (\sspds) are indispensable in fields such as quantum science and technology, astronomy, and biomedical imaging, where high detection efficiency, low dark count rates and high timing accuracy are required.
Recently, helium (He) ion irradiation was shown to be a promising method to enhance \sspd performance.
Here, we study how changes in the underlying superconducting \nbtin film and the \siox/Si substrate affect device performance.
While irradiated and unirradiated \nbtin films show similar crystallinity, we observe He bubble formation below the \siox/Si interface and an amorphization of the Si substrate.
Both reduce the \csublong from \qty{210}{\W\per\m^2\K^4} to \qty{70}{\W\per\m^2\K^4} after irradiation with \qty{2000}{\ipsn}.
This effect, combined with the lateral straggle of He ions in the substrate, allows the modification of the superconductor-to-substrate \csub of an \sspd by selectively irradiating the regions around the nanowire.
With this approach, we achieved an increased plateau width of saturating intrinsic detection efficiency of \qty{9.8}{\uA} compared to \qty{3.7}{\uA} after full irradiation.
Moreover, the critical current remained similar to that of the unirradiated reference device (\qty{59.0}{\uA} versus \qty{60.1}{\uA}), while full irradiation reduced it to \qty{22.4}{\uA}.
Our results suggest that the irradiation-induced reduction of the \csub significantly enhances \sspd sensitivity, offering a novel approach to locally engineer substrate properties for improved detector performance.

\end{abstract}

\maketitle

\section{Introduction}
Superconducting nanowire single-photon detectors (\sspds) \cite{Goltsman2001} play an increasing role in a multitude of fields such as quantum science and technology \cite{Takesue2007, Chen2022, Liu2023, Bussieres2014, Takesue2015, Shibata2014, Valivarthi2016, Natarajan2012}, optical communication \cite{Grein2015, Biswas2017}, astronomy \cite{Wollman2021}, biology \cite{Xia2021, Wang2022, Tamimi2024}, and medicine \cite{Ozana2021, Poon2022, Parfentyeva2023}.
With their near-unity efficiency \cite{Marsili2012a, Korneev2012}, sub-\unit{\Hz} dark count rate \cite{Shibata2015}, picosecond timing jitter \cite{Korzh2020}, and nanosecond reset time \cite{Cherednichenko2021}, they are ideally suited for demanding applications such as photonic quantum computing \cite{Slussarenko2019, Gyger2021, Ferrari2018, Sprengers2011, Reithmaier2013, Reithmaier2015, Majety2023}, lidar \cite{Chen2013, Korzh2020, Guan2022, Kaifler2022},
particle and dark matter detection \cite{Pena2024, Polakovic2020, Shigefuji2023, Hochberg2019, Chiles2022},
or infrared fluorescence microscopy for in vivo deep brain imaging \cite{Xia2021, Wang2022, Tamimi2024}.
To further advance photon detection efficiency, lower dark count rate, as well as improving timing jitter and the homogeneity of device performance in \sspd arrays, helium (He) ion irradiation has emerged as a valuable post-processing technique \cite{Zhang2019, Strohauer2023, Charaev2024, Batson2024, Strohauer2024}.
Previous studies have found that He ion irradiation alters properties of the underlying superconducting films. 
For example, it increases the sheet resistance and decreases the critical temperature of NbN, \nbtin, and \ce{MgB2} superconducting thin films due to the formation of defects \cite{Zhang2019, Strohauer2023, Charaev2024, Batson2024}.  %
Since the critical temperature is directly proportional to the superconducting energy gap, this results in an enhanced sensitivity of \sspds \cite{Zhang2019}.
Moreover, Charaev \etal \cite{Charaev2024} note that for their \ce{MgB2} based single-photon detectors, He ion irradiation not only introduces defects but also amorphizes the underlying SiC substrate, likely affecting the \csublong. This leads to a prolonged lifetime of the normal conducting domain, thereby increasing the detection probability.
However, it has remained unclear whether such mechanisms, in addition to irradiation-induced defect formation, play a significant role in the improved performance after irradiation \cite{Charaev2024, Batson2024}. 
In this work, we investigate how He ion irradiation modifies the morphology of both the \nbtin film and the \siox/Si substrate and discuss mechanisms by which these changes influence detector performance.

\section{Irradiation-induced changes of film and substrate morphology}
\label{sec:irradiation-induced-changes-of-film-and-substrate}

\begin{figure*}
 \centering
 \includegraphics[width=\textwidth]{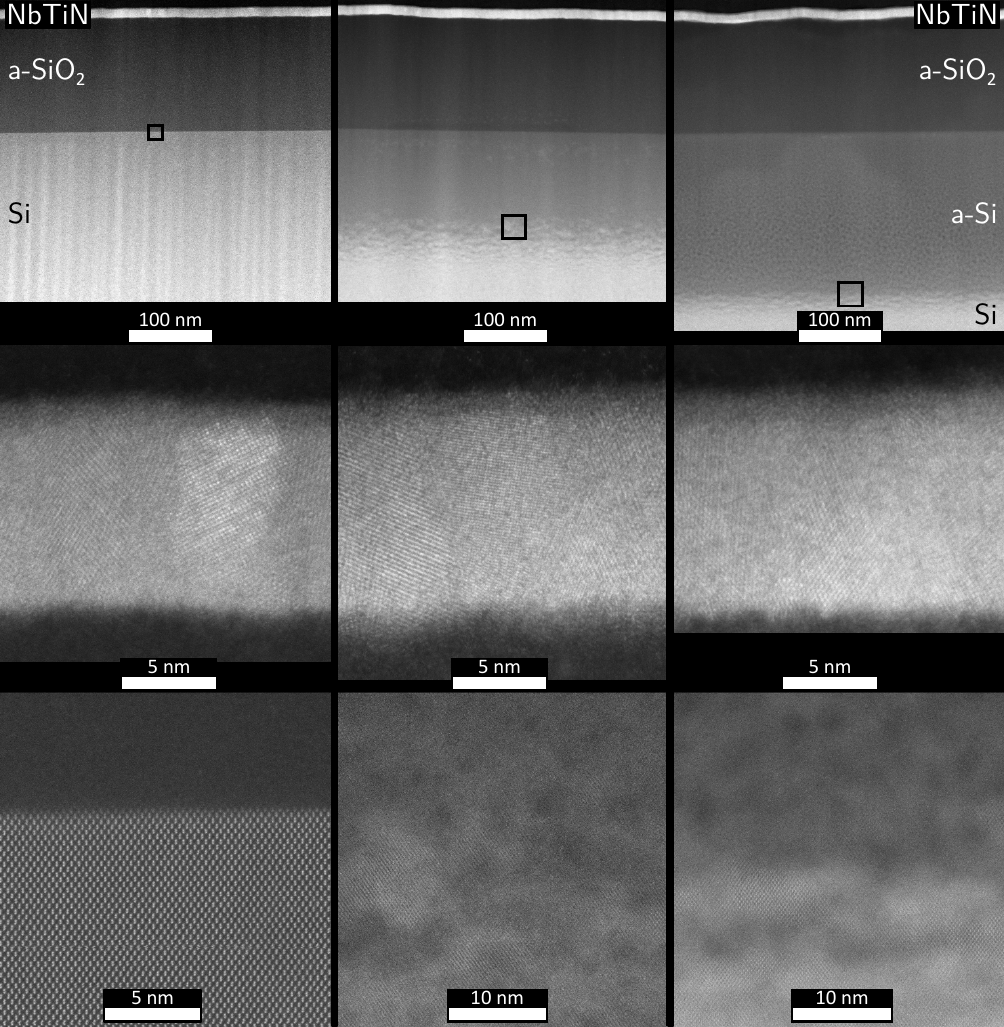}
     \caption{
     HAADF-STEM images of \nbtin/\siox/Si stacks before and after He ion irradiation: unirradiated (left column), irradiated with \qty{500}{\ipsn} (middle column), and \qty{2000}{\ipsn} (right column). 
     Top row: Overview images showing the \nbtin film (light gray), the amorphous \siox (dark gray), and the Si substrate. 
     In the irradiated samples, the upper Si regions become amorphous (a-Si) with grayscale variations indicating porosity. 
     The crystalline Si appears light gray.
     Middle row: High-resolution images of the \nbtin films reveal a similar thickness of \qty{11.3}{\nm}, \qty{5}{\nm}-wide columnar $\delta$-\nbtin domains, and a \qty{1.5}{\nm} thick amorphous oxygen-rich surface layer across all samples. 
     Bottom row: Transition regions between amorphous and crystalline Si (locations marked by black rectangles in the top row). 
     The crystalline-amorphous interface coincides with the \siox/Si interface in the unirradiated sample, but lies \qty{105}{\nm} and \qty{190}{\nm} below the \siox/Si interface for the samples irradiated with \qty{500}{\ipsn} and \qty{2000}{\ipsn}, respectively. 
     Dark contrasts in the amorphous Si regions are associated with pores.
     }
 \label{fig:Composition_TEM_pictures}
\end{figure*}

To explore how He ion irradiation influences superconducting film and substrate properties, we fabricated nominally \qty{12}{\nm} thick \nbtin films via DC reactive magnetron sputtering on \siox/Si substrates with a nominally \qty{130}{\nm} thick thermally grown \siox layer.
Subsequently, we irradiated selected areas with various fluences of \qty{30}{\kV} He ions before characterizing them using a transmission electron microscope (TEM) and an atomic force microscope (AFM).
\cref{fig:Composition_TEM_pictures} shows 
high-angle annular dark-field scanning transmission electron microscopy (HAADF-STEM) images of the stack
\nbtin/\siox/Si in its unirradiated form, and after irradiation with \qty{500}{\ipsn} and \qty{2000}{\ipsn} (corresponding to \qty{5e16}{\ions\per\cm\squared} and  \qty{2e17}{\ions\per\cm\squared}).
Following He ion irradiation, we observe that the Si substrate becomes amorphous and contains He bubbles in a region extending from 
the \siox/Si interface at \qty{150}{\nm} depth to 
approximately \qty{350}{\nm} depth, consistent with previous studies \cite{Raineri2000, Livengood2009, Abrams2012, Hlawacek2014, Allen2021}.

\begin{figure*}
 \centering
 \includegraphics{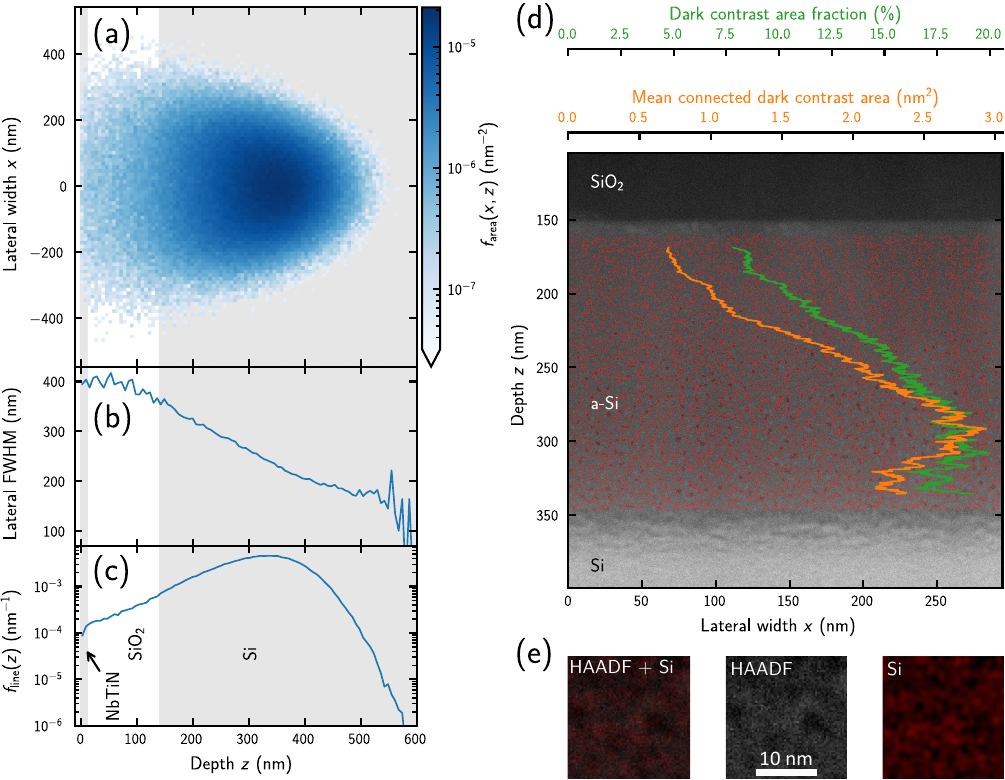}
     \caption{
     Simulated distribution of He ion stopping positions (a-c) and experimental analysis of structural modifications (d-e) in the irradiated \nbtin/\siox/Si stack.
     (a) Lateral and depth resolved probability density function $f_\mathrm{area}(x, z)$ of the stopping positions. 
     (b) Depth resolved FWHM of the lateral Gaussian distribution. (c) Probability density function $f_\mathrm{line}(z)$ for He ions stopping at a certain depth $z$. 
     (d) HAADF-STEM image of the sample irradiated with \qty{2000}{\ipsn}, with dark contrast regions highlighted by red outlines. Overlaid curves show the dark contrast area fraction (green) and average size of connected dark contrast regions (orange) versus depth $z$.  
     (e) High-magnification comparison of HAADF contrast with Si content:
     overlay of HAADF and Si EDX signal (left), HAADF image (center), and Si EDX map (right), demonstrating the correlation between dark HAADF contrast and reduced Si content.
     }
     \label{fig:Composition_SRIM_and_TEMporeAnalysis}
\end{figure*}

This finding agrees with Monte Carlo simulations for the same \nbtin/\siox/Si stack, performed with the SRIM 2008 software \cite{Ziegler2010}.
\cref{fig:Composition_SRIM_and_TEMporeAnalysis}a-c shows the calculated lateral and depth resolved probability density function $f_\mathrm{area}(x, z)$ of stopping positions of the He ions originating from the focused beam of a He ion microscope, as well as the depth resolved FWHM of the lateral Gaussian distribution, and the probability density function $f_\mathrm{line}(z)$ of He ions stopping at a certain depth $z$.
The lateral distribution was calculated by projecting the three-dimensional stopping positions of He ions onto a plane perpendicular to the surface.
Most of the He ions penetrate deep into the substrate before they stop, with the maximum of the distribution of implanted He ions at a depth of \qty{335}{\nm}, which is in the range where we observed He bubble formation.
At the same time, the He distribution is broadened laterally and in depth due to scattering and recoil events.
Therefore, the simulated lateral Gaussian distribution has a FWHM of \qty{266}{\nm} when considering all He ion stopping positions, \qty{241}{\nm} for the layer where most He ions stop, and \qty{400}{\nm} within the \nbtin layer (although the fraction of ions that stop within each nm of this layer is more than an order of magnitude smaller than deep in the substrate).

Furthermore, the HAADF-STEM images in \cref{fig:Composition_TEM_pictures} reveal that He ion irradiation with \qty{500}{\ipsn} (\qty{2000}{\ipsn}) transforms crystalline Si into an amorphous and porous structure with He bubbles.
This extends to a depth of \qty{105}{\nm} (\qty{190}{\nm}) below the \siox/Si interface, corresponding to a total depth of \qty{255}{\nm} (\qty{340}{\nm}).
Interestingly, the Si layer near the \siox/Si interface remains partially crystalline after irradiation with \qty{500}{\ipsn} (see \cref{sec:crystalline_SiO2-Si_interface}), while for the higher fluence of \qty{2000}{\ipsn}, the amorphous region extends all the way to the \siox-Si interface.
The thickness of the \siox layer, however, remains unchanged at \qty{137+-3}{\nm} after irradiation.
As shown in \cref{fig:Composition_SRIM_and_TEMporeAnalysis}e, complementary HAADF-STEM imaging and energy-dispersive X-ray spectroscopy (EDX) of the amorphous Si region reveal that areas of dark contrast correspond to reduced Si content.
These dark contrasts, originating from local minima in the projected Si thickness, are associated with He bubble formation during irradiation. 
This interpretation is supported by the irradiation-induced surface elevation of the stack measured by AFM, as discussed later in this section.
To quantify these observations, we analyzed the spatial distribution of dark contrast in the amorphous Si region in \cref{fig:Composition_SRIM_and_TEMporeAnalysis}d. 
At each depth, we calculated the dark contrast area fraction by identifying regions below a threshold value after local background subtraction. 
This analysis shows that both the area fraction and average size of dark contrast regions reach their maxima at approximately \qty{300}{\nm} depth, closely matching the simulated peak of He ion stopping positions at \qty{335}{\nm} shown in \cref{fig:Composition_SRIM_and_TEMporeAnalysis}c.
We observe these dark contrast regions in a range extending from approximately the \siox/Si interface (located at \qty{150}{\nm} depth) to about \qty{350}{\nm} depth. 
While the projective nature of TEM imaging prevents direct interpretation of individual feature sizes, this spatial correlation between dark contrast regions and simulated ion stopping positions, combined with the observed surface elevation in AFM measurements, indicates the formation of He bubbles in the Si substrate, consistent with previous studies \cite{Raineri2000, Livengood2009, Abrams2012, Hlawacek2014, Allen2021}.
Moreover, the \nbtin film exhibits increased wrinkling at higher He ion fluences, while it maintains its original thickness of \qty{11.3 +- 0.5}{\nm} and shows no visible cracks or noticeable material loss from sputtering.
We use the term "wrinkling" rather than "surface roughness" because the film undergoes a characteristic deformation where both its top and bottom surfaces exhibit synchronized undulations while maintaining constant thickness.
This wrinkling and its correlation with AFM surface roughness measurements are analyzed in detail in \cref{sec:nbtin_film_wrinkling_vs_roughness}. 
Notably, despite the wrinkling of the \nbtin layer, the \siox/Si interface remains flat.
Furthermore, at all fluences the \nbtin film retains its columnar crystalline domains  (NaCl-type face centered cubic $\delta$-phase of \nbtin) with an oxygen-rich top layer (see \cref{sec:EDX} for the elemental composition) \cite{Oya1974, Brauer1952, Yamamoto1991}.
We do not see any substantial change in crystallinity or visible defects between the differently irradiated \nbtin films.
This is surprising, since we expected the polycrystalline structure of the pristine \nbtin film to become more amorphous and exhibit an increasing number of defects with increasing He ion fluence.

\begin{figure*}
 \centering
 \includegraphics{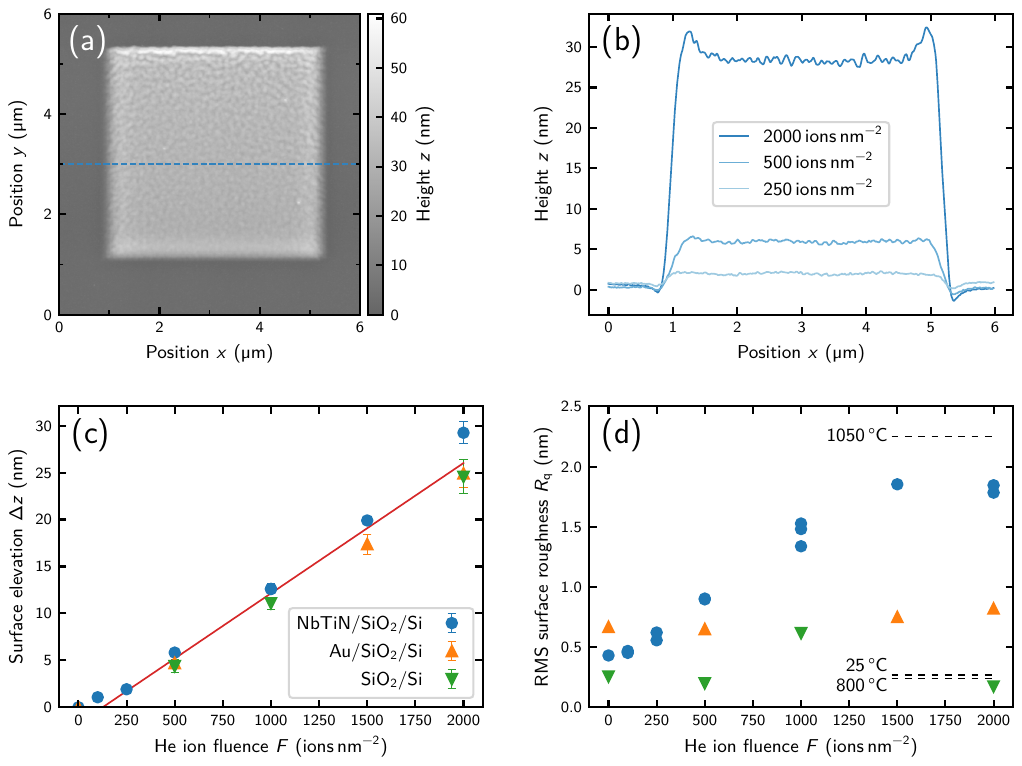}
     \caption{AFM characterization of irradiated \nbtin films. (a) A representative square structure obtained after irradiation with \qty{2000}{\ipsn}. (b) Line profiles for three different He ion fluences. For \qty{2000}{\ipsn}, the profile is also indicated as a blue dashed line in (a). (c) The fluence dependent surface elevation of the quasi-flat center of the irradiated area. Besides the stack \nbtin/\siox/Si, it was also measured for the reference stacks Au/\siox/Si and \siox/Si. The red line is a linear fit of all data with a fluence of \qty{250}{\ipsn} or above. (d) Dependence of the RMS surface roughness on the He ion fluence for the same stacks as in (c). The dashed lines show the measured roughness before and after annealing a second \nbtin/\siox/Si sample at temperatures of \qty{800}{\celsius} and \qty{1050}{\celsius} (see \cref{sec:baking} for details).
     }
 \label{fig:composition_AFM}
\end{figure*}

Additionally to the TEM characterization, the irradiated surfaces are investigated using an AFM (Bruker Dimension Icon) to gain information about the surface morphology and to compare it to the TEM measurements.
\cref{fig:composition_AFM}a shows the surface of an irradiated \qtyproduct[product-units = repeat]{4 x 4}{\um} \ce{NbTiN} square. 
Upon irradiation with $\qty{2000}{\ipsn}$, we observe a surface elevation of \qty{28}{\nm} and an increased surface roughness within the irradiated area. 
The elevation predominantly stems from He bubble formation in the silicon, as the density difference between amorphous and crystalline silicon is only \qty{1.8}{\percent} \cite{Custer1994}.
During irradiation, the scan direction causes uneven redeposition of sputtered \nbtin, with more material accumulating at the starting edge (upper edge). 
This results in an increased height of that edge compared to the others \cite{Yoon2017, Bachmann2020}.
To quantify the dependence of the surface elevation and the surface roughness on the He ion fluence, we performed AFM measurements for multiple fluences and not only for the \nbtin/\siox/Si stack but also for two reference stacks, Au/\siox/Si and \siox/Si.
\cref{fig:composition_AFM}b shows the height profiles of three differently irradiated \nbtin/\siox/Si stacks, where each point represents the average height across a \qty{1.5}{\um} wide strip perpendicular to the profile direction.
As shown in \cref{fig:composition_AFM}c, the surface elevation increases linearly with the He ion fluence for both Au and \nbtin thin films, as well as for the bare \siox/Si substrate. 
This behavior is consistent with expectations, since most He ions interact only weakly with the film and stop deep inside the substrate where they form He bubbles that cause the surface to rise.
The linear fit (excluding fluences below \qty{250}{\ipsn}) has a non-zero intercept at \qty{120}{\ipsn}, from which we conclude that significant bubble formation and surface elevation begin just above a certain fluence, similar to an observation in \cite{Kajita2011}.
To study the dependence of the root mean square (RMS) roughness of irradiated \nbtin films on the He ion fluence, we irradiated squares with dimensions of \qtyproduct[product-units = repeat]{4x4}{\um} and \qtyproduct[product-units = repeat]{15 x 15}{\um} with beam currents ranging from \qtyrange{10.6}{119}{\pico\ampere}, which were chosen to obtain irradiation times between \qty{1}{\min} and \qty{10}{\min} even for the highest fluences. 
As shown in \cref{fig:composition_AFM}d, the surface roughness of the \ce{NbTiN} film  
increases from \qty{0.43}{\nm} for the unirradiated film to \qty{1.85}{\nm} at \qty{1500}{\ipsn}. 
On the contrary, the bare \siox/Si substrate and the sample with a \qty{10}{\nm} gold layer do not show a systematic increase in roughness after irradiation (\qty{0.31}{\nm} and \qty{0.72}{\nm} average roughness, respectively). 
Moreover, while the different squares of \qtyproduct[product-units = repeat]{4x4}{\um} and \qtyproduct[product-units = repeat]{15 x 15}{\um} exhibit similar roughness in the homogeneously irradiated center regions, we observed a slightly higher surface roughness for higher beam currents as discussed in more detail in \cref{sec:appendix_beam_current}.
Furthermore, the surface roughness observed in the AFM measurements of this section correlates with the wrinkling identified in the TEM images above.
This relationship is analyzed in detail in \cref{sec:nbtin_film_wrinkling_vs_roughness}.

Measurements in the literature have shown that ion irradiation enables plastic flow in \siox and Si 
\cite{Abrams2012, Snoeks1994, Volkert1989, Volkert1993, Volkert1991}, and that sputtered \nbtin typically exhibits internal stress \cite{Iosad1999, Cecil2007, Yamamori2010, Shan2015, Karimi2016}.
This agrees with our observation of an increasing wrinkling  of the \nbtin film with increasing He ion fluence:
When irradiating the \nbtin/\siox/Si stack, the stress in the \nbtin film can relax and deform the originally flat \nbtin film due to radiation-induced plastic flow of the underlying \siox.
To test this hypothesis, we took a \nbtin film on a \siox/Si substrate of the same sputtering run and heated it to temperatures up to \qty{1050}{\celsius}.
Since viscous flow of thermally grown \siox starts at \qty{960}{\celsius} \cite{EerNisse1977}, we expected an increase of the \nbtin film wrinkling just above that temperature.
Indeed, AFM line profiles do not differ between the original sample and the sample after heating it to \qty{800}{\celsius}.
However, after heating the sample to \qty{1050}{\celsius}, it shows a substantial increase of wrinkling with the RMS roughness increasing from \qty{0.27}{\nm} to \qty{2.25}{\nm} (more details in \cref{sec:baking}).
This observation indicates that stress in the \nbtin film is already present after sputtering and that it relaxes upon activating viscous flow of the underlying substrate.
To verify that the deformation does not originate from strain in the substrate material, we also obtained reference data by irradiating the unprocessed wafer (\siox/Si stack) and the wafer with an evaporated gold layer on top.
Both do not show an increase of wrinkling/roughness with increasing He ion fluence, as can be seen in \cref{fig:composition_AFM}d.

In summary, we observe an increased wrinkling of the \nbtin film as the He ion fluence increasess, while its thickness and roughness remain unchanged.
This wrinkling originates from relaxation of stress that builds up in the \nbtin film already during the sputtering process.
When the \nbtin/\siox/Si stack is irradiated with He ions or heated to temperatures above \qty{960}{\celsius}, viscous flow of the \siox allows stress in the \nbtin layer to be released, causing the \nbtin to wrinkle.

\section{Impact of He ion irradiation on the \csub}

We continue to link the He-ion induced changes in the superconducting thin film and the substrate to their thermal coupling, before showing in the subsequent section that this contributes substantially to the irradiation-induced performance enhancement of \sspds.
The detection efficiency in SNSPDs is influenced not only by thin film properties such as critical temperature, sheet resistance, and diffusivity, but also by thermal properties of the thin film and the superconductor-substrate interface. 
This is highlighted by the improved detection efficiency that Xu \etal achieved after reducing the \csublong by underetching the \sspd \cite{Xu2023a}. 
Simulations in this work showed that samples with lower \csub exhibit a longer hotspot decay time, meaning that the heat loss from the hotspot to the substrate is suppressed, thereby increasing the fraction of the photon energy that contributes to hotspot formation.
We expect that the effects of He ion irradiation, such as the formation of He bubbles and the modification of the crystallinity of the Si substrate, may similarly reduce the \csub and play an important role in the observed increase in detection efficiency of \sspds following irradiation.
Therefore, we continue by investigating the impact of He ion irradiation on the \csublong in the following.

\begin{figure*}
 \centering
 \includegraphics{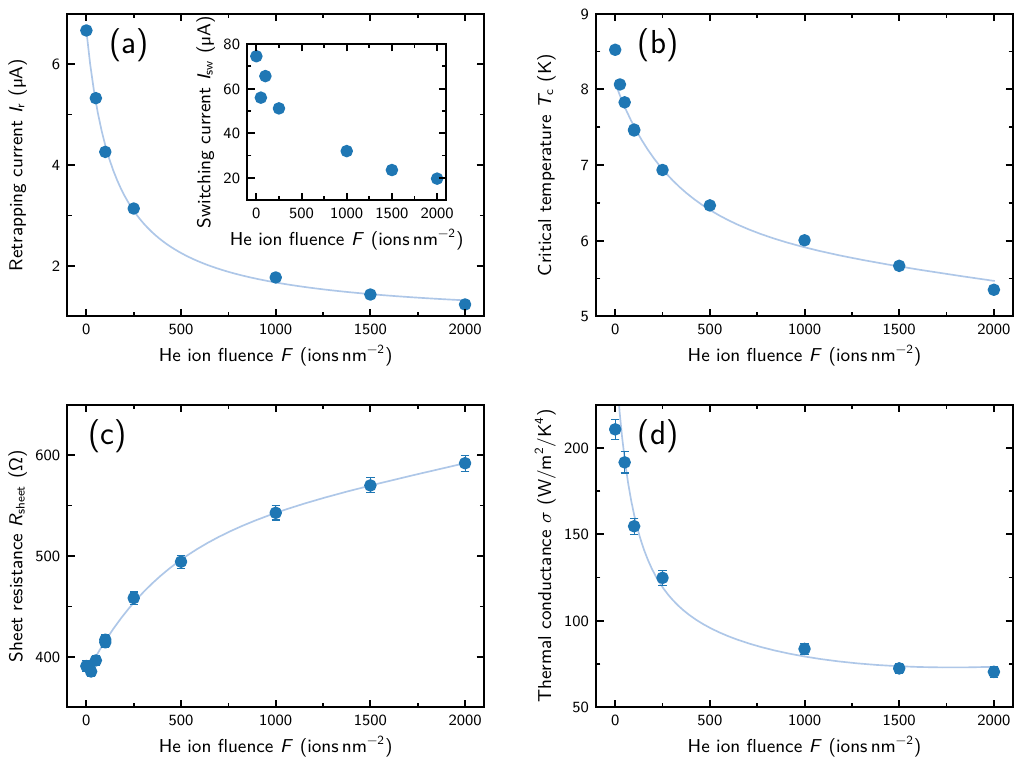}
     \caption{
     Modification of basic transport properties and of the \csub from the \sspd to the substrate by He ion irradiation. (a) The retrapping current $I_\mathrm{r}$ with the switching current $\Isw$ as inset, both measured at a temperature of \qty{1}{\K}. (b) The critical temperature $\Tc$. (c) The sheet resistance $\Rsheet$. (d) The \csublong $\sigma$. While the curves for $I_\mathrm{r}$ and $\sigma$ are solely guides to the eye, the curves for $\Tc$ and $\Rsheet$ were fitted with the models described in the main text. Error bars are in some cases smaller than the symbols.
     }\label{fig:composition_heatConductance_transport_currents}
\end{figure*}

The equation governing heat flow in a one-dimensional nanowire is
\begin{equation}
    J^2\rho + \kappa \frac{\partial^2 T}{\partial x^2} - \frac{\sigma}{d}(T^4-T_\mathrm{sub}^4) = C \frac{\partial T}{\partial t},
\end{equation}
accounting for the heat flow along the nanowire, Joule heating from the resistive barrier when the nanowire is in the normal conducting state, and heat escaping from the superconducting thin film of thickness $d$ to the substrate at a fixed substrate temperature $T_\mathrm{sub}$ \cite{Stockhausen2012a,Sidorova2022,Xu2023a,Xu2021a}.
Here, $J$ is the current density, $\rho$ is the normal-state resistivity, $\kappa$ is the thermal conductivity of the  superconducting nanowire, $\sigma$ is the \csublong, and $C$ is the specific heat per unit volume of NbTiN. 
According to the literature \cite{Dane2022, Xu2023}, the \csub $\sigma$ can be obtained from the equation for the retrapping current density,
\begin{equation}
    J_\mathrm{r} = \sqrt{\frac{\sigma}{4d\rho}
    \left(
    T_\mathrm{c}^4-T_\mathrm{sub}^4
    \right)
    } \;,\label{eq:heat_conductance_hyst_current}
\end{equation}
with the critical temperature $\Tc$.
To measure the retrapping current $I_\mathrm{r}$, 
we pass a current above the switching current $I_\mathrm{sw}$ through the nanowire and then gradually reduce it until the normal conducting resistive domain within the nanowire can no longer sustain itself by Joule heating. 
This current, at which it switches back to the superconducting state, defines the retrapping current. 
We measured $I_\mathrm{r}$ for \qty{25.8}{\um} long, \qty{250}{\nm} wide, and \qty{8}{\nm} thick nanowires at a temperature of \qty{1}{\K}, irradiated with He ion fluences ranging from \qtyrange{0}{2000}{\ipsn}. 
As shown in \cref{fig:composition_heatConductance_transport_currents}a, the largest decrease of $I_\mathrm{r}$ occurs at fluences smaller than \qty{250}{\ipsn} and it reduced from \qty{6.7}{\uA} to \qty{1.2}{\uA} comparing the unirradiated device to the device after irradiation with \qty{2000}{\ipsn}.
Moreover, in the inset of \cref{fig:composition_heatConductance_transport_currents}a we present the measured switching current of the same devices and observed the expected decrease with increasing He ion fluence that has already been reported in the literature \cite{Zhang2019, Strohauer2023}.
In addition to the retrapping and the switching current, we measured the critical temperature and the sheet resistance of the irradiated films using the van der Pauw method as described in previous work \cite{vanderPauw1958, Miccoli2015, Strohauer2023}.
As shown in \cref{fig:composition_heatConductance_transport_currents}b and c, the critical temperature decreases, while the sheet resistance increases with increasing He ion fluence.
We fitted the sheet resistance using a model of our previous work \cite{Strohauer2023} that accounts for defect formation in the \nbtin film, and relate the critical temperature to the ion fluence via the sheet resistance by using the universal scaling law of Ivry \etal \cite{Ivry2014}. 
While the sheet resistance fit closely matches the measured data, the critical temperature fit differs slightly from measurements at low and high He ion fluences.
With these ingredients and the substrate temperature $T_\mathrm{sub} = \qty{1}{\K}$, we calculate the \csub $\sigma$ via \cref{eq:heat_conductance_hyst_current}.
As shown in \cref{fig:composition_heatConductance_transport_currents}d, it decreased from initially \qty{210}{\watt\per\m^2\kelvin^4} to \qty{70}{\watt\per\m^2\kelvin^4}
after irradiation with \qty{2000}{\ipsn}.
Moreover, the data suggests a saturating \csub for fluences above \qty{1500}{\ipsn}.
This can be explained by considering two paths contributing to the total \csub: (1) a path where heat escapes normal to the wafer surface through the irradiated area, $\sigma_\mathrm{norm}$, and (2) a path where heat flows in-plane, parallel to the wafer surface, $\sigma_\textnormal{in-plane}$.
Since the He bubbles form deep within the substrate and the He ions deposit most of their energy there, we expect that He ion irradiation primarily reduces $\sigma_\textnormal{norm}$ deep inside the substrate and underneath the irradiated area.
At the same time, we expect $\sigma_\textnormal{in-plane}$ in the \nbtin film and in the first nanometers underneath the film to be less affected by the irradiation.
Above a sufficiently high He ion fluence, the in-plane \csub starts to dominate and the total thermal conductance saturates at a value determined by the lateral dimensions of the irradiated area and the thickness of the thermally conducting channel.
The \csub values obtained in this work lie slightly below the range reported in literature for NbN (\qtyrange{229}{711}{\watt\per\m^2\kelvin^4}) \cite{Xu2021a,Xu2023a,Dane2022}, which could be attributed to the different material properties of \nbtin versus NbN.

In summary, the \csublong decreases continuously with increasing He ion fluence, and after irradiation with \qty{2000}{\ipsn} it saturates at only \qty{33}{\percent} of the value for the unirradiated device. 
The reduced \csub can have multiple origins such as (1) the He bubbles that form deep inside the substrate and act as a thermally insulating layer, (2) the irradiation-induced amorphization of the originally crystalline Si \cite{Zink2006,Thompson1961}, and (3) a modified superconductor-substrate interface due to the relaxation of internal stress in the \nbtin.

\section{Enhancing detection efficiency by engineering the \csub}
\label{sec:enhancing_DE_by_heat_conductance_engineering}
In the previous sections we elaborated that the most prominent irradiation-induced changes of the \nbtin film and the underlying \siox/Si substrate are the wrinkling of the \nbtin film, the amorphization of the first few hundred nanometers of Si, 
and the formation of He bubbles in a region extending from approximately the \siox/Si interface (located at \qty{150}{\nm} depth) to about \qty{350}{\nm} depth.
Besides the fact that point defects are not visible in our TEM images of the \nbtin films, we also did not observe any changes of crystallinity or amorphization of the \nbtin film with increasing He ion fluence.
Therefore, we conclude that in addition to the direct impact of irradiating the \nbtin film, a significant contribution to the enhanced sensitivity of \sspds following He ion irradiation \cite{Zhang2019, Strohauer2023, Strohauer2024} arises from a change in the \csublong.

To verify that changes in detector performance after irradiation cannot solely be explained by the direct irradiation effects of the \nbtin film,
we compare three different irradiation schemes: 
(1) an unirradiated reference device, (2) a device fully irradiated with \qty{2000}{\ipsn}, and (3) a device where we irradiated only the surrounding regions up to a distance of \qty{150}{\nm} to the superconducting wire with \qty{2000}{\ipsn}.
The focused He ion beam enters the substrate, where through scattering and recoil events, the He ions successively transfer their energy to the substrate while spreading laterally as they move through it.
Due to this lateral straggle, we expect the He ions to amorphize the Si and create He bubbles deep inside the substrate not only in the directly irradiated regions, but also underneath the nanowire for the device where we irradiated only the regions next to the superconducting wire.
This will modify the \csub to the substrate, while keeping the direct effects of He ions on the \nbtin wire limited.
\cref{app:He_ion_distribution_at_distant_irradiation} contains a detailed discussion about the simulated He ion distribution and energy deposition, and about the limitations of this particular irradiation scheme.
We conclude that if a direct modification of the \nbtin wire by the He ion beam was the one and only cause for a change of detector properties, an irradiation of the surrounding area should have a very limited influence on them.

\begin{figure*}
 \centering
 \includegraphics{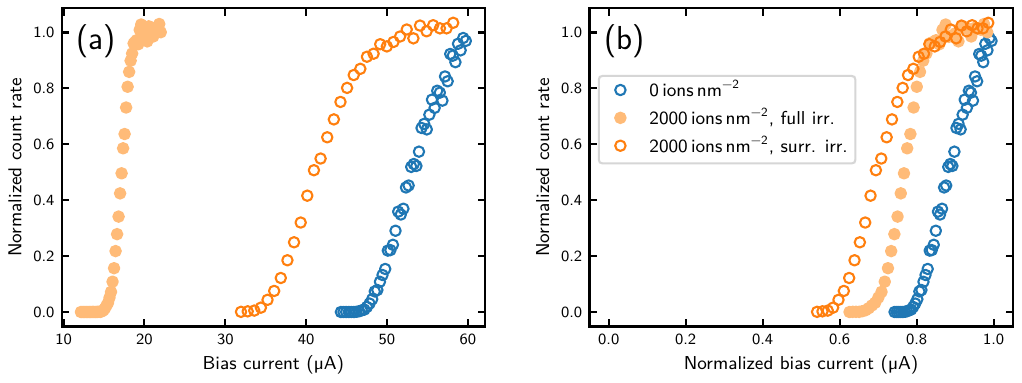}
     \caption{
     Normalized count rate versus bias current for three differently irradiated \sspds.
     (a) The count rate is normalized to the fit asymptote after fitting an error function, and plotted against the absolute bias current. (b) The same data as in (a) but with the bias current normalized to the critical current of the \sspd. The measurements were performed at a wavelength of \qty{780}{\nm} and a temperature of \qty{1}{\K}.
     }
 \label{fig:composition_CR_surr_vs_full_irr}
\end{figure*}

As shown in \cref{fig:composition_CR_surr_vs_full_irr}a and b, our experiments with \sspds consisting of \qty{250}{\nm} wide and \qty{8}{\nm} thick nanowires at a temperature of \qty{1}{\K} demonstrate
that both irradiated devices exhibit saturating detection efficiency at a wavelength of \qty{780}{\nm}, while the unirradiated device does not reach saturation.
Moreover, we observe that while the critical current \footnote{The critical current is defined as the smallest current at which the \sspd shows a finite resistance, being either in the latching or in the relaxation oscillation regime. The switching current, on the other hand, is defined as the current where the \sspd switches to the normal conducting state. 
Note also that we measured the critical current with a \qty{10.2}{\ohm} shunt resistor connected during the count rate measurements, while the switching current was measured without a shunt resistor and without illumination.} 
of the fully irradiated device is reduced to \qty{22.4}{\uA}, the device where only the surrounding area was irradiated exhibits a similar critical current as the unirradiated device (\qty{59.0}{\uA} and \qty{60.1}{\uA}, respectively).
At the same time, the switching current is not only reduced for the fully irradiated device but also for the device where only the surrounding area was irradiated.
The subtleties when comparing the critical currents and the switching currents of those devices are elaborated in detail in \cref{app:comparison_Isw_vs_Ic}.
It is noteworthy that the relative saturation plateau width is of similar size (\qty{16.6}{\percent}) and the absolute saturation plateau width is substantially larger for the wire with only the surrounding area irradiated 
compared to the fully irradiated wire (\qty{9.8}{\uA} compared to \qty{3.7}{\uA}).
This is particularly interesting since the wire itself was not irradiated and, moreover, the number of He atoms underneath the wire is on average only \qty{3.9}{\percent} of that underneath the fully irradiated wire (see \cref{app:He_ion_distribution_at_distant_irradiation}).
These observations imply that irradiating only the surrounding area increases the sensitivity due to the reduced \csub to the substrate, and simultaneously the sensitivity stays high since the critical current is barely reduced.
In contrast, we typically observe a strong reduction of the critical current after full irradiation of \sspds \cite{Strohauer2023}.
This also means that for devices where only the surrounding area is irradiated, much higher He ion fluences could be used to obtain even larger saturation plateau widths.

The most important conclusion of this experiment is that despite irradiating only the surrounding region of the nanowire, its sensitivity to single photons increased significantly.
This would not be possible if a modification of the \nbtin via direct He ion irradiation was the only cause for a change in detector properties after irradiation.
Thus, other mechanisms such as the irradiation-induced reduction of the \csub from the detector to the substrate
substantially contribute to the change in detector properties after irradiation.

\section{Conclusion}

Motivated by the significant improvement of \sspd performance following He ion irradiation \cite{Zhang2019, Strohauer2023, Strohauer2024}, we investigated the irradiation-induced changes of the superconducting \nbtin films and the underlying \siox/Si substrates.
Through combined TEM and AFM analysis, we found that irradiating the \nbtin/\siox/Si stack with \qty{30}{\kV} He ions transforms the crystalline Si into an amorphous structure, with He bubbles forming between the \siox/Si interface (at \qty{150}{\nm} depth) and \qty{350}{\nm} depth, causing substrate swelling.
At the same time, the \nbtin film remains polycrystalline with columnar crystalline domains and an oxygen-rich top layer.
However, the originally flat \nbtin film exhibits an increasing wrinkling with increasing He ion fluence since stress in the as-sputtered \nbtin film can be released due to viscous flow of the \siox that allows its plastic deformation during He ion irradiation.

Moreover, experiments revealed a decrease in the \csub from the \nbtin to the substrate with increasing He ion fluence from initially \qty{210}{\W\per\m^2\K^4} to \qty{70}{\W\per\m^2\K^4} after irradiation with \qty{2000}{\ipsn}.
Combined with the observations for the influence of He ion irradiation on film and substrate morphology, this suggests the following: 
in addition to the effects of direct He ion irradiation on the \nbtin film, a substantial contribution to the enhanced \sspd performance originates from an irradiation-induced reduction of the \csub.
This reduction, caused by He bubbles, amorphization of crystalline Si, and a modified \nbtin-\siox interface, leads to a prolonged lifetime of the normal-conducting domain, thereby increasing the detection probability.
To verify this hypothesis, we irradiated the surrounding area of a straight nanowire up to a distance of \qty{150}{\nm} from the wire.
Due to scattering, the He ions spread laterally when moving through the sample and result in a reduced \csub also underneath the nanowire.
Although we did not irradiate the nanowire directly, we observed a similar relative saturation plateau width (\qty{16.6}{\percent}), a significantly larger absolute saturation plateau width (\qty{9.8}{\uA} compared to \qty{3.7}{\uA}), and a higher critical current (\qty{59.0}{\uA} compared to \qty{22.4}{\uA}) after irradiation of the surrounding area of the nanowire with \qty{2000}{\ipsn}, compared to full irradiation with the same He ion fluence.
This confirms that the reduced \csub substantially contributes to the enhanced sensitivity following He ion irradiation.
Moreover, irradiation of the area surrounding the nanowire provides a means to reduce the \csub underneath the nanowire while maintaining a high critical current due to the limited interaction of He ions with the \nbtin film.

To further differentiate the effects of the He ions on the \nbtin film versus the substrate, it would be desirable to compare detectors fabricated on pre-irradiated substrate regions with those fabricated on unirradiated substrate regions and then irradiated.
Moreover, since the He implantation depth, which also defines the thickness of the heat conducting channel, can be tuned via the acceleration voltage, investigating the influence of different acceleration voltages on both device and substrate properties could provide valuable insights for optimizing detector performance.

\section*{Acknowledgments}
The authors highly appreciate FIB sample preparation by Dr.~Steffen Schmidt.
We gratefully acknowledge support from the German Federal Ministry of Education and Research (BMBF) via the funding program ``Photonics Research Germany'' (projects MOQUA (13N14846) and MARQUAND (BN106022)) and the funding program ``Quantum technologies -- from basic research to market'' (projects PhotonQ (13N15760), SPINNING (13N16214), QPIS (16K1SQ033), QPIC-1 (13N15855) and SEQT (13N15982)), as well as from the German Research Foundation (DFG) under Germany's Excellence Strategy EXC-2111 (390814868) with the projects PQET (INST 95/1654-1) and MQCL (INST 95/1720-1).
This research is part of the ``Munich Quantum Valley'', which is supported by the Bavarian state government with funds from the ``Hightech Agenda Bayern Plus''.

\section*{Author Declarations}
\subsection*{Conflict of interest}
The authors have no conflicts to disclose.

\section*{Data availability}
The data that support the findings of this study are openly available in Harvard Dataverse at \url{https://doi.org/10.7910/DVN/QQ31HM}.

\appendix

\section{Fabrication and characterization of \nbtin films and \sspds}
To fabricate the superconducting \nbtin films and \sspds studied in this work, we deposited \nbtin films using DC reactive magnetron sputtering onto two \ce{Si} substrates (\qtyproduct[product-units = repeat]{5x5}{\mm}) with a nominally \qty{130}{\nm} thick thermally grown \siox layer.
The superconductor thickness was controlled by measuring the sputtering rate and choosing the sputtering time correspondingly.
On the sample that was later used for AFM and TEM measurements, we deposited a nominally \qty{12}{\nm} thick \nbtin layer. 
Subsequent TEM images revealed an actual thickness of \qty{11.3+-0.5}{\nm}. 
On the other sample we deposited a nominally \qty{8}{\nm} thick \nbtin layer for fabricating \sspds in the form of straight nanowires of \qty{250}{\nm} width and \qty{25.8}{\um} length to study the influence of He ion irradiation onto the \csublong and an associated change in detection efficiency.
On the nanowire sample, we patterned the \nbtin film with electron beam lithography and reactive ion etching, followed by contact pad fabrication using optical lithography and gold evaporation \cite{Flaschmann2023}.
The process for contact pad fabrication was also used to fabricate gold marker structures on the sample for AFM and TEM measurements.

After characterization of the unirradiated devices, we used a He ion microscope (Zeiss Orion Nanofab) with an acceleration voltage of \qty{30}{\kV} for irradiation with He ions.
On the nanowire sample, we fabricated three types of devices:
(1) unirradiated nanowires, (2) fully irradiated nanowires, and (3)
nanowires for which only the surrounding area was irradiated, leaving the nanowire and a \qty{150}{\nm} wide stripe next to the nanowire unirradiated.
On the AFM/TEM sample, we used the He ion microscope to irradiate \qtyproduct[product-units= repeat]{4x4}{\um} and \qtyproduct[product-units= repeat]{15x15}{\um} squares of the \nbtin film.
Subsequently, we measured those irradiated areas with an AFM (Bruker Dimension Icon) using the Scan Asyst mode and an OTESPA tip.

Cross-sectional TEM lamellae were prepared from three regions: an unirradiated square and squares irradiated with \qty{500}{\ipsn} and \qty{2000}{\ipsn}. 
Using a focused ion beam microscope (FEI Helios G3 UC), we thinned these sections to approximately \qtyrange{10}{35}{\nm} in the vicinity of the NbTiN film. 
The lamellae were characterized using a probe-corrected transmission electron microscope (FEI Titan Themis) operated at \qty{300}{\kV}, employing (scanning) transmission electron microscopy ((S)TEM), high-angle annular dark-field imaging (HAADF), and energy-dispersive X-ray spectroscopy (EDX). 
All images presented in \cref{fig:Composition_TEM_pictures} were acquired using HAADF-STEM.

\section{Partial retention of Si crystallinity near the \siox/Si interface after irradiation with \qty{500}{\ipsn}
}
\label{sec:crystalline_SiO2-Si_interface}

\begin{figure}
 \centering
 \includegraphics[width=\columnwidth]{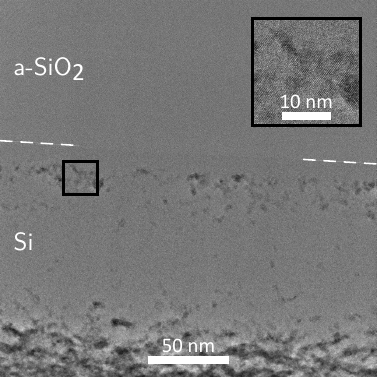}
     \caption{
     TEM image of the interfacial region between amorphous \siox and Si in the \nbtin/\siox/Si stack after irradiation with \qty{500}{\ipsn}. The interface is marked by a dashed white line. Light gray regions indicate amorphous material and darker areas reveal crystalline silicon.
     The inset ($3\times$ magnification) shows a partially crystalline region where lattice fringes are clearly visible, demonstrating the local crystalline order.
     }
 \label{fig:Si-SiO2_interface_details}
\end{figure}

After irradiation with \qty{500}{\ipsn}, the Si substrate exhibits a distinct transition region near the \siox interface where crystalline and amorphous phases coexist, as shown in \cref{fig:Si-SiO2_interface_details}. 
This partial preservation of crystallinity stands in contrast to the complete amorphization observed at \qty{2000}{\ipsn}, indicating that the transformation process depends strongly on the He ion fluence. 
The high-resolution inset in \cref{fig:Si-SiO2_interface_details} reveals clear lattice fringes in the preserved crystalline domains, providing direct evidence of residual crystalline order in these regions.

\section{\nbtin film wrinkling and surface roughness}
\label{sec:nbtin_film_wrinkling_vs_roughness}
\begin{figure}
 \centering
 \includegraphics{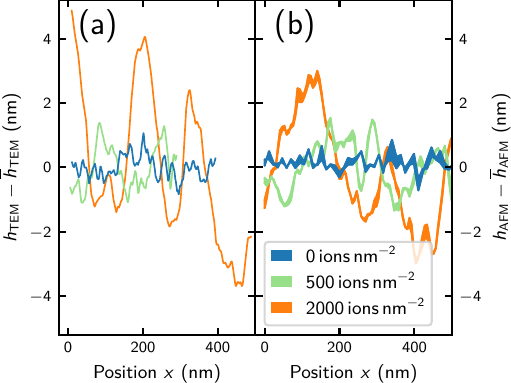}
     \caption{
     Local deviation from the surface height average, $h - \overline{h}$, as determined by (a) TEM and (b) AFM measurements of the unirradiated \nbtin film and after irradiation with \qty{500}{\ipsn} and \qty{2000}{\ipsn}. 
     The local width of the lines in (b) is defined by the local height range bounded by the AFM line profiles obtained from the two averaging widths (\qty{12}{\nm} and \qty{35}{\nm}).}
     \label{fig:composition_TEM_vs_AFM_wrinkliness}
\end{figure}

The TEM measurements presented in \cref{sec:irradiation-induced-changes-of-film-and-substrate} reveal that the \nbtin film maintains its thickness after irradiation while exhibiting increasing wrinkling, characterized by synchronized undulations of its top and bottom surfaces. 
To quantify this wrinkling and compare it with AFM measurements, \cref{fig:composition_TEM_vs_AFM_wrinkliness}a shows the local deviation from the mean surface height of the \nbtin layer as determined from the TEM images. 
The analysis reveals that the local surface height varies within intervals of \qtylist{1.8; 2.5; 8}{\nm} around the mean surface height for the film before He ion irradiation and after exposure to \qtylist{500;2000}{\ipsn}, respectively.
To compare this wrinkling with the AFM-measured surface roughness presented in \cref{sec:irradiation-induced-changes-of-film-and-substrate}, \cref{fig:composition_TEM_vs_AFM_wrinkliness}b shows line profiles from three AFM images for the same He ion fluences.
Each height value along these profiles is the local average across a strip oriented perpendicular to the profile direction, using widths of \qty{12}{\nm} and \qty{35}{\nm} to match the thickness range of the TEM lamellae. 
The local width of the lines in \cref{fig:composition_TEM_vs_AFM_wrinkliness}b is defined by the local height range, bounded by the profiles obtained from the two averaging widths, and indicates that both yield similar results.
The AFM line profiles exhibit surface height variations within intervals of \qtylist{1.6; 3; 6}{\nm} around the mean surface height for the unirradiated film and after exposure to \qtylist{500;2000}{\ipsn}, respectively. 
These values, as well as their characteristic length scales of variation, closely match those derived from the TEM images, demonstrating quantitative agreement between both measurement techniques.
Based on these results, we conclude that the AFM line profiles and RMS roughness presented here and in \cref{sec:irradiation-induced-changes-of-film-and-substrate} not only show continuously increasing surface height variations with increasing He ion fluence, but can also be used to estimate the wrinkling of the \nbtin film observed in TEM images.
To conclude, the increase of the AFM-measured surface roughness with increasing He ion fluence, combined with the correlation between surface roughness and wrinkling, indicate that the \nbtin film undergoes wrinkling rather than developing surface roughness with increasing He ion fluence.

\section{Energy-dispersive X-ray spectroscopy}
\label{sec:EDX}
\begin{figure*}
 \centering
 \includegraphics{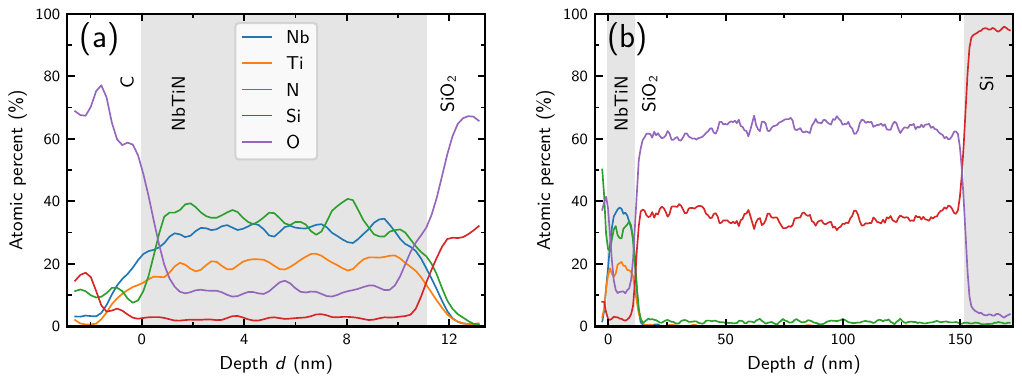}
     \caption{
     Energy dispersive X-ray spectroscopy (EDX) line profiles extracted from STEM-EDX maps of the unirradiated \nbtin/\siox/Si stack. 
     The profiles were taken perpendicular to the interfaces, with integration performed parallel to them.
     (a) High-resolution scan across the NbTiN film (\qty{20}{\nm} integration width), showing the \siox support (right), the \nbtin film with its oxygen-rich surface layer (center), and a protective carbon layer required for sample preparation (left). 
     Carbon was excluded from the elemental quantification to prevent distortion from its residual presence throughout the sample. 
     (b) Wide-area scan across the complete \nbtin/\siox/Si stack (\qty{58}{\nm} integration width).
}
 \label{fig:Composition_EDX_linescans}
\end{figure*}

For compositional analysis of the \nbtin film and the \nbtin/\siox/Si stack, we performed energy-dispersive X-ray spectroscopy (EDX) using scanning transmission electron microscopy (STEM). 
From these EDX maps, we extracted integrated compositional profiles perpendicular to the interfaces, as shown in \cref{fig:Composition_EDX_linescans}.

\section{Influence of the annealing temperature on the surface roughness}
\label{sec:baking}
\begin{figure*}
 \centering
 \includegraphics{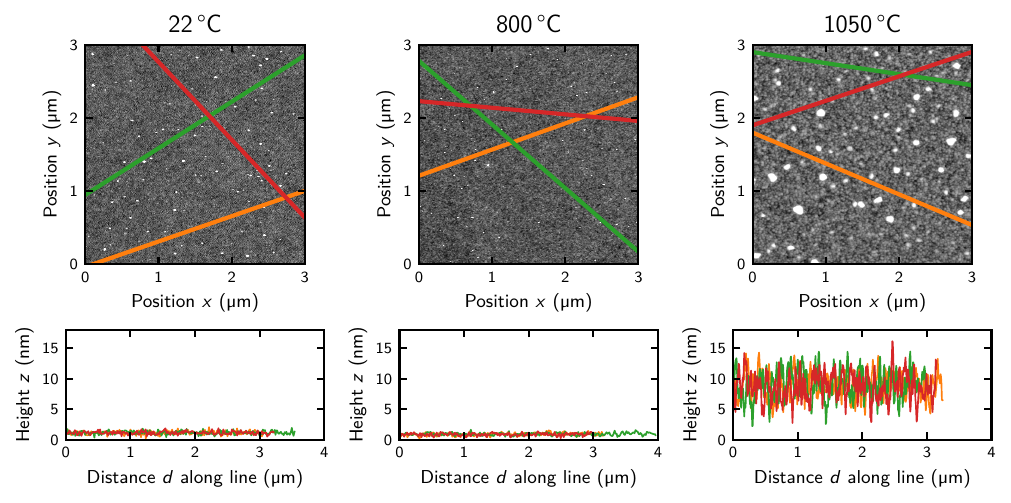}
     \caption{AFM measurements and corresponding line profiles of the stack \nbtin/\siox/Si after exposing it to three different temperatures. The as-sputtered film and the film after heating it to \qty{800}{\celsius} exhibit a similar RMS roughness of \qty{0.27}{\nm} and \qty{0.24}{\nm}, respectively. This increases to \qty{2.25}{\nm} after heating the sample to \qty{1050}{\celsius}.
     }
 \label{fig:baking_NbTiN_SiO2_Si_AFM_results}
\end{figure*}

In this section we investigate in how far stress in the originally flat \nbtin film can relax and deform the \nbtin film by heating the sample to temperatures above \qty{960}{\celsius} where viscous flow of thermally grown \siox starts \cite{EerNisse1977}.
To this aim, we mounted a sample with the stack \nbtin/\siox/Si in a chamber that was purged with nitrogen and evacuated to a pressure range of \qtyrange{e-7}{e-8}{\milli\bar} before ramping up the temperature and holding the target temperature for \qty{30}{\min}.
\cref{fig:baking_NbTiN_SiO2_Si_AFM_results} shows the AFM images and line profiles across the surface before and after heating to \qty{800}{\celsius} and \qty{1050}{\celsius}.
We note that the sample used in this heating experiment initially had many large grains (potentially from a non-clean surface before sputtering) that grew in size when heating to \qty{1050}{\celsius}. 
Therefore, we placed the line profiles in regions without those grains.
While the line profiles before and after heating the sample to \qty{800}{\degree} show a similar RMS roughness of \qty{0.27}{\nm} and \qty{0.24}{\nm}, it increases to \qty{2.25}{\nm} after heating to \qty{1050}{\celsius}.
This indicates that stress in the \nbtin film is already present after sputtering and that it relaxes when enabling viscous flow of the underlying substrate.

\section{Influence of the beam current on the surface roughness}
\label{sec:appendix_beam_current}
\begin{figure}
 \centering
 \includegraphics{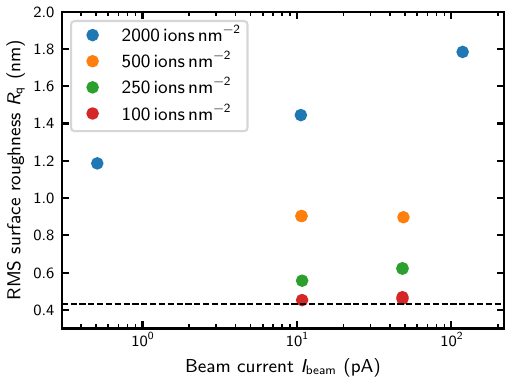}
     \caption{RMS surface roughness versus beam current that was used for irradiation with He ion fluences ranging from \qtyrange{100}{2000}{\ipsn}. The black dashed line indicates the RMS surface roughness before irradiation.
     }
 \label{fig:RMS_surface_roughness_vs_beam_current}
\end{figure}

To investigate the influence of the beam current on the roughness/wrinkling of the \nbtin film, we irradiated the stack \nbtin/\siox/Si with beam currents ranging from \qtyrange{0.5}{118}{\pico\ampere}.
\cref{fig:RMS_surface_roughness_vs_beam_current} shows the RMS surface roughness after irradiation as measured with an AFM.
The roughness increases from \qty{0.43}{\nm} before irradiation to \qty{1.19}{\nm} and \qty{1.78}{\nm} for the lowest and highest beam current, respectively.
This difference we explain by an increased viscous flow of the \siox at higher beam currents, allowing more relaxation of the stress that built up in the \nbtin during the sputtering process.
However, even for the smallest beam currents, the roughness increased by a factor of three after irradiation with \qty{2000}{\ipsn} compared to the unirradiated film.

\section{He ion distribution for distant irradiation}
\label{app:He_ion_distribution_at_distant_irradiation}
\begin{figure*}
 \centering
 \includegraphics{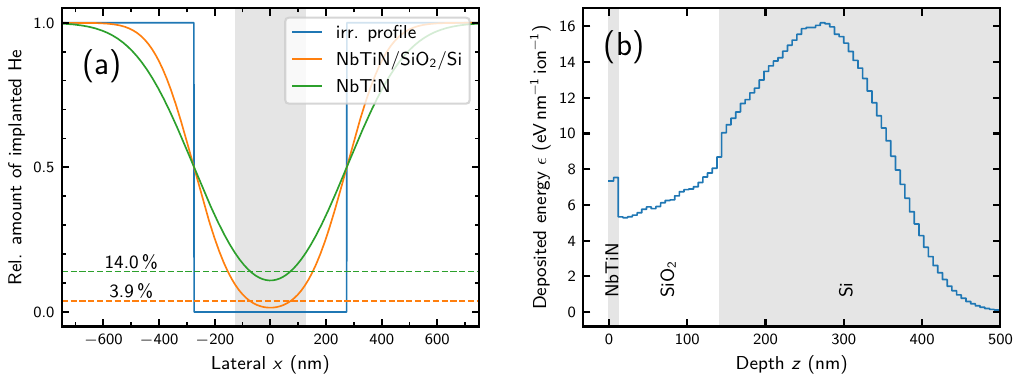}
     \caption{Simulated He ion and energy distribution after irradiation of the sample stack \nbtin/\siox/Si in the surrounding area of the wire up to a distance of \qty{150}{\nm} from the wire. (a) Normalized implanted He ion distribution. The irradiation profile is shown as a blue line (1 in the regions where the He ion beam hits the surface), while the width of the nanowire is indicated by the gray shaded area. The orange line is the distribution of He ions integrated along the full depth of the sample (over the whole stack \nbtin/\siox/Si), while the green line is the distribution when integrating only over the \nbtin layer. Both distributions are normalized to the amount of implanted He for full irradiation. The corresponding average amount of implanted He underneath the wire is indicated by the horizontal dashed lines. (b) Deposited energy per ion and per nm sample depth. The material stack \nbtin/\siox/Si is indicated by the differently shaded areas.}
 \label{fig:composition_SRIM_outerIrr_energyDistribution}
\end{figure*}

In this section we estimate the relative amount of implanted He ions and deposited energy in and underneath a \qty{250}{\nm} wide \nbtin wire when irradiating only the surrounding area up to a distance of \qty{150}{\nm} to the wire, and compare it to a fully irradiated wire.
The average lateral spread of He ions when irradiating a single spot can be calculated by projecting the simulated three-dimensional stopping positions of He ions onto a line parallel to the surface, and results in a Gaussian distribution with a FWHM of \qty{266}{\nm}.
When irradiating only the regions next to the superconducting wire, this lateral spread will result in a significant fraction of He ions underneath the nanowire.
The relative fraction of implanted He ions $f_\mathrm{rel}$ along a line perpendicular to the nanowire can be calculated by taking the convolution of the Gaussian distribution of He ions with the one-dimensional profile of irradiation along this line,
\begin{equation}
    f_\mathrm{rel}(x)
    =
    \int_{-\infty}^{\infty}
    \frac{1}{\sqrt{2\pi\sigma^2}}e^{-\frac{(s-x)^2}{2\sigma^2}}
    \cdot
    \Theta\left(\left| s \right| - \frac{w_\mathrm{unirr}}{2}\right)
    \dif s
    \;,
\end{equation}
with the standard deviation $\sigma$ of the Gaussian distribution, the width $w_\mathrm{unirr}$ of the unirradiated area, and the Heaviside step function $\Theta$.
The resulting He ion distribution is shown as orange line in \cref{fig:composition_SRIM_outerIrr_energyDistribution}a, with the minimum and maximum amount of He ions underneath the nanowire being \qty{1.5}{\percent} and \qty{9.3}{\percent}, and an average amount of \qty{3.9}{\percent} of that in the fully irradiated area.

As shown in \cref{fig:Composition_SRIM_and_TEMporeAnalysis}b, the FWHM of the He ion stopping positions within the \nbtin layer is \qty{400}{\nm} due to recoil events and scattering.
Following the same procedure as above but projecting only the stopping positions of He ions that end up in the \qty{12}{\nm} thick \nbtin film onto a line perpendicular to the
nanowire,
we calculate the convolution of the resulting Gaussian distribution with the one-dimensional irradiation profile and obtain the relative fraction of He ions implanted in the nanowire as presented as green line in \cref{fig:composition_SRIM_outerIrr_energyDistribution}a.
Since the FWHM of the implanted He distribution in the \nbtin layer is higher than that when considering the stopping positions of all He ions, also the minimum and maximum relative fractions, \qty{10.9}{\percent} and \qty{20.1}{\percent}, and the average fraction of \qty{14.0}{\percent} are higher.
This also means that the number of He ions that stop within the \nbtin wire is in the same order of magnitude when comparing full irradiation with irradiation up to a distance of \qty{150}{\nm} to the wire.
This leads to the question how the effects of direct irradiation differ from irradiation up to a distance of \qty{150}{\nm} to the wire, since according to the Bragg peak the He ions deposit most energy per nm shortly before they stop, as can be seen in \cref{fig:composition_SRIM_outerIrr_energyDistribution}b.
However, the energy loss per nm when entering the sample is still half of that at the maximum in \qty{270}{\nm} sample depth.
Moreover, at the spots where the focused He ion beam hits the sample, all He ions deposit a fraction of their kinetic energy in the \nbtin film.
Taking into account that only the fraction \num{e-4} of all ions end up in each nm of the \nbtin film, we conclude that the total energy deposited in the \nbtin wire when irradiating up to a distance of \qty{150}{\nm} to the wire, is approximately four orders of magnitude smaller than when irradiating the whole device.
We note, however, that the small fraction \num{e-4} of the He ions that are implanted in the \nbtin wire per nm is similar when comparing fully irradiated devices with those irradiated up to a distance of \qty{150}{\nm} to the wire.
At the same time, the total amount of He ions implanted in the \siox/Si substrate underneath the nanowire is still \qty{3.9}{\percent} of that in the fully irradiated area, leading to substantial He bubble formation and a reduced superconductor-to-substrate \csub underneath the nanowire.

Thus, when keeping the wire unirradiated and irradiating only its surroundings up to a distance of \qty{150}{\nm} to the wire, the \nbtin should be barely directly affected by the He ion beam (only by the small fraction of deflected or recoiled atoms/ions), while the reduced \csub due to the amorphization of Si and the formation of He bubbles in the substrate extends under the nanowire.

We note that beside the amorphization of Si and the formation of He bubbles underneath the wire, we cannot completely exclude that the following two points also have some influence on the device performance:
(1) The curved surface due to the different surface elevation in the center versus the edges of the nanowire, which is originating from smaller He bubbles underneath the center of the wire compared to underneath the wire edges.
(2) A potential residual plastic flow of the \siox near the \nbtin interface due to the small amount of energy deposited in this region from scattered/recoiled He ions. 
However, since the fraction of energy is only on the order of \num{e-4} compared to the fully irradiated regions, we expect this effect to be negligible.
To fully isolate the change in device properties due to the modified superconductor-to-substrate \csub after irradiation, future studies comparing the performance of devices fabricated on pre-irradiated substrate regions to similar devices fabricated on unirradiated regions of the same substrate would be useful.

\section{Comparison of critical current and switching current}
\label{app:comparison_Isw_vs_Ic}

We note that the critical current $\Ic$ determined from the count rate measurements with a shunt resistor (\qty{10.2}{\ohm} at \qty{1}{\K}) differs from the switching current $\Isw$ measured without shunt resistor as shown in \cref{tab:Isw_vs_Ic}.
The critical current is defined as the smallest current at which the \sspd shows a finite resistance, being either in the latching or in the relaxation oscillation regime. The switching current, on the other hand, is defined as the current where the \sspd switches to the normal conducting state. 
The fact that the switching current is lower than the critical current for the devices with the full wire or the surrounding area irradiated 
might be explained by
latching, since for switching current measurements no shunt resistor was connected.
Surprisingly, the switching current for the unirradiated device is higher than its critical current.
However, when comparing the two devices ``surrounding irradiated'' and ``fully irradiated'', a clear trend can be observed:
Despite that the switching current for both devices is reduced after irradiation, the reduction is much less pronounced for the device with the surrounding irradiated 
(\qty{42.5}{\uA} compared to \qty{19.7}{\uA}, corresponding to reductions by \qty{47}{\percent} and \qty{79}{\percent}).
This is even more apparent when comparing the critical current after and the switching current before irradiation (a reduction by \qty{27}{\percent} compared to \qty{76}{\percent}).
The fact that also the switching current of the device where only the surrounding area was irradiated is reduced might be explained by He ions that reach the nanowire due to back scattering and recoil events as elaborated in \cref{app:He_ion_distribution_at_distant_irradiation}.

We conclude that even if the unirradiated device had shown a critical current similar or higher than its switching currents, the arguments in \cref{sec:enhancing_DE_by_heat_conductance_engineering} remain the same:
The relative saturation plateau width of the device for which the surrounding area was irradiated shows a similar width as that of the fully irradiated device, while its switching current and its critical current remain significantly higher.

\begin{table}
    \centering
    \begin{tabular}{c!{\qquad}S!{\qquad}S!{\qquad}S}
    \toprule
                            &  {unirr.}         & {surr.  irr.}  & {fully irr.} \\
    \midrule
 $\Isw$ before irradiation  &  79.1             & 80.8                      & 95.2 \\
 $\Isw$ after irradiation   &  74.5             & 42.5                      & 19.7 \\
 $\Ic$ after irradiation    &  60.1             & 59.0                      & 22.4 \\
    \bottomrule
    \end{tabular}
    \caption{Comparison of critical currents and switching currents (in  \unit{\uA}) of the three devices discussed in \cref{sec:enhancing_DE_by_heat_conductance_engineering}. The unirradiated device was always measured as reference device and also in the rows ``$\Isw$ after irradiation'' as well as ``$\Ic$ after irradiation'' it was never irradiated.}
    \label{tab:Isw_vs_Ic}
\end{table}

\bibliography{HIM_paper_III}%

\end{document}